\documentclass[a4paper,11pt]{article}
\usepackage{latexsym}
\usepackage{epsfig, subfigure, subfloat, graphicx, float }
\usepackage{amssymb}
\usepackage{epstopdf}
\author{H. Mohseni Sadjadi \footnote{mohsenisad@ut.ac.ir} and Parviz Goodarzi
\\ {\small Department of Physics, University of Tehran,}
\\ {\small P. O. B. 14395-547, Tehran 14399-55961, Iran}}
\title{Reheating in non-minimal derivative coupling model}

\begin{document}
\maketitle
\begin{abstract}
We consider a model with non-minimal derivative coupling of
inflaton to gravity.  The reheating process during rapid
oscillation of the inflaton is studied and the reheating
temperature is obtained. Behaviors of the inflaton and produced
radiation in this era are discussed.
\end{abstract}

\section{Introduction}
To solve some problems of standard model of cosmology and particle
physics, such as the horizon problem,  flatness, isotropy and
homogeneity of the universe, the absence of magnetic monopoles and
so on, and to make the bing bang cosmology more consistent with
astrophysical data, the inflation theory, which considers an epoch
of accelerated expansion for the early universe, was introduced
\cite{guth}. One straightforward way to describe this era, is to
consider a slowly rolling scalar field , dubbed inflaton, whose
energy density was dominated by its potential during inflation
\cite{inflaton1,inflaton2}. At the end of inflation the universe
was cold, so there must be a procedure trough which the scalar
field decayed to particles which became thermalized and reheated
the universe. This could be realized by decaying of the scalar
field, e.g. during coherent oscillation in the bottom of the
potential \cite{reheating}.

At first sight, it seems that a candidate for this scalar field
may be the Higgs boson. But parameters of the standard model do
not agree with those required for inflaton \cite{inflaton2}. So to
reconciliate the slow roll inflation with standard model
parameters, a framework in which the Higgs boson is non minimally
coupled to Ricci scalar has been introduced in \cite{berz}.

Recently a model comprising a non minimal coupling between the
derivatives of the Higgs boson and Einstein tensor has been
proposed, which besides its capacity to explain the inflationary
phase, is also safe of quantum corrections and unitary violation
problem \cite{germani}. In this context, the non-minimal
derivative coupling may allow the model to describe the
acceleration as well as the super-acceleration of the universe
\cite{sad}. Coupling the Einstein tensor to kinetic term of
inflaton, as we will see later, enhances the gravitational
friction during slow-roll and this allows us to consider more
general steep potentials, such as Higgs potential, without
contradiction with the CMB observations or collider experimental
bound \cite{germani}.

In \cite{kehagias}, this model was employed to study the natural
inflation, where the inflaton is assumed to be a
pseudo-Nambu-Goldstone boson. In this framework, the global shift
symmetry is broken at a scale $f$, giving rise to the inflaton
mass. For small field values, the potential is stable against
radiative corrections, but slow roll requires that $f$ becomes
much larger than the Planck scale, giving rise to eta problem. In
\cite{kehagias}, it was shown that non-minimal derivative coupling
allows to take $f\ll M_P$, without introducing new degrees of
freedom, and protects the tree-level shift invariance of the
scalar field as well as the perturbative aspects of the theory.

Similar models including nonminial derivative coupling between a
scalar field and gravity have also been used to study the late
time evolution of the universe by considering the scalar field as
the dark energy \cite{late}.

In this manuscript, following \cite{germani} we assume that the
inflation is implemented by a scalar field with non-minimal
derivative coupling to gravity, with a power law potential, and
study the reheating process in this model, which to our knowledge,
was not studied before. First, we review briefly the inflationary
epoch and then study quasi-periodic motion of the inflaton at the
end of the slow roll. We consider the decay of the scalar field to
ultra-relativistic particles (radiation) via a phenomenological
source during coherent rapid oscillation and find the reheating
temperature.

\section{Non-minimal derivative coupling model and inflation}

An action describing a scalar field coupled non-minimally to
gravity via its kinetic term is given by\footnote{We use units
$\hbar=c=1$ though the paper.} \cite{germani}:
\begin{equation}\label{1}
S=\int \Big({M_P^2R\over 2}-{1\over 2}g^{\mu \nu}\partial_\mu
\varphi \partial_{\nu} \varphi +{w\over 2}G^{\mu \nu}\partial_\mu
\varphi \partial_\nu \varphi- V(\varphi)\Big)\sqrt{-g}d^4x,
\end{equation}
where $G^{\mu \nu}$ is the Einstein tensor, $w$ is a constant with
the dimension of inverse mass squared, and $M_P=2.4\times
10^{18}GeV$ is the reduced Planck mass. In the absence of terms
containing more than two time derivatives, additional degrees of
freedom are not produced in this theory.

In the spatially flat Friedmann-Lemaitre-Robertson-Walker (FLRW)
space-time, the Friedmann equation is
\begin{equation}\label{2}
H^2={1\over 6M_P^2}\Big((1+9wH^2)\dot{\varphi}^2+2V(\varphi)\Big),
\end{equation}
and the scalar field equation of motion is given by
\begin{equation}\label{3}
(1+3wH^2)\ddot{\varphi}+3H(1+3wH^2+2w\dot{H})\dot{\varphi}+V'(\varphi)=0,
\end{equation}
where "dot" denotes derivative with respect to time and "prime"
denotes derivative with respect to $\varphi$.

Using the energy momentum tensor derived from (\ref{1}), the
energy density and the pressure of the scalar field are obtained
as
\begin{eqnarray}\label{4}
\rho_{\varphi}&=&(1+9wH^2){\dot{\varphi}^2\over
2}+V(\varphi),\nonumber
\\
P_{\varphi}&=&{\dot{\varphi}^2\over 2}-V(\varphi)-{w\over
2}(3H^2+2\dot{H})\dot{\varphi}^2-2wH\dot{\varphi}\ddot{\varphi},
\end{eqnarray}
respectively. $\rho_{\varphi}$ and $P_{\varphi}$ satisfy the continuity equation
\begin{equation}\label{z}
\dot{\rho_{\varphi}}+3H(P_{\varphi}+\rho_{\varphi})=0.
\end{equation}

In the presence of another component, with the energy density
$\rho_{\mathcal{R}}$ and the pressure $P_{\mathcal{R}}$,
interacting with the scalar field via the source term $Q$, the
continuity equation becomes
\begin{eqnarray}\label{34}
&&\dot{\rho_{\varphi}}+3H(P_{\varphi}+\rho_{\varphi})=-Q, \nonumber \\
&&\dot{\rho_{\mathcal{R}}}+3H(P_{\mathcal{R}}+\rho_{\mathcal{R}})=Q.
\end{eqnarray}
In the inflationary era, we take the field $\varphi$ as the
inflaton, and assume that the universe is dominated by only this
scalar field. But in the subsequent epochs, one must also take
into account the presence of other components such as radiation.

In the following we adopt the high friction condition \cite{hf}
\begin{equation}\label{f}
wH^2\gg 1.
\end{equation}
The slow roll regime is characterized by
\begin{eqnarray}\label{5}
&&\ddot{\varphi}\ll 3H\dot{\varphi},\hspace{1cm} {|\dot{H}|\over
H^2}\ll 1,\nonumber \\
&&9wH^2\dot{\varphi}^2\ll 2V(\varphi),
\end{eqnarray}
yielding
\begin{eqnarray}
H^2&\simeq& {V(\varphi)\over 3M_P^2}\nonumber \\
\dot{\varphi}&\simeq& -{V'(\varphi)\over 9wH^3}
\end{eqnarray}
The slow-roll conditions are satisfied provided that
\begin{eqnarray}\label{6}
&&{V''(\varphi)\over V^2(\varphi)} \ll {3w\over M_P^4}, \nonumber
\\
&&{{V'}^2(\varphi)\over V^3(\varphi)} \ll {2w\over M_P^4}.
\end{eqnarray}
In comparison with minimal models, conditions (\ref{6}) may be
satisfied by more steep potentials in the high friction limit
(\ref{f}) \cite{hf}.

For the potential
\begin{equation}\label{N1}
V(\varphi)=\lambda \varphi^q ,
\end{equation} these relations
require $\varphi^{q+2}\gg {M_P^4\over w\lambda}$. In the slow-roll
era, where $\dot{H}\ll H^2$ holds, the requiring that the model be
outside of the quantum gravity regime implies that $R\simeq
12H^2\ll {M_P^2\over 2}$. This condition, in terms of the
potential, is rewritten as
\begin{equation}\label{7}
V(\varphi)\ll {M_P^4\over 8}.
\end{equation}
The number of e-folds is given by
\begin{equation}\label{8}
\mathcal{N}=\int_{\varphi_{e}}^{\varphi_0}Hdt={w\over
M_P^4}\int_{\varphi_{e}}^{\varphi_0} {V^2(\varphi)\over
V'(\varphi)}d\varphi,
\end{equation}
where $\varphi_{e}$ $(\varphi_{0})$ is the value of the scalar
field at the end (the beginning) of the slow-roll inflation.

By assuming $\varphi_{e}\ll \varphi_{0}$, one can estimate the
e-folds number for a chaotic inflation with potential (\ref{N1})
as
\begin{equation}\label{9}
\mathcal{N}\simeq {w\lambda\over q(q+2)M_P^4}\varphi_{0}^{q+2}.
\end{equation}
When the conditions (\ref{6}) cease to be valid
\begin{eqnarray}\label{Y1}
&&{V''(\varphi)\over V^2(\varphi)} \simeq {3w\over M_P^4},
\nonumber
\\
&&{{V'}^2(\varphi)\over V^3(\varphi)}\simeq {2w\over M_P^4},
\end{eqnarray}
the slow-roll inflation ends. For power law potential (\ref{N1})
and $q\sim \mathcal{O}(1)$, (\ref{Y1}) gives
\begin{equation}\label{m}
\varphi_{e}^{q+2}\simeq {q^2 M_P^4\over 3w \lambda}.
\end{equation}

\section{Quasi-periodic evolution}

\begin{figure}[h!]
\centering\epsfig{file=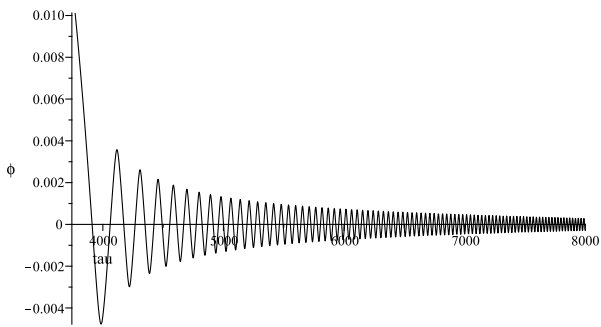,width=8cm}
\caption{${\phi:={\varphi\over M_P}}$ in terms of dimensionless
time $\tau=mt$, for $\{wm^2=10^{8} $
with initial conditions $\{\phi(1)=0.056$, $\dot{\phi}(1)=0\}$, for the quadratic potential. }
\label{fig1}
\end{figure}

In this part we try to study the dynamics of the inflaton after
the end of slow-roll. In this era, conditions (\ref{5},\ref{6})
cease to be valid and quasi-periodic evolution of the scalar field
about the bottom of the potential begins. In fig.(\ref{fig1}),
using numerical methods, oscillation of the field $\varphi$ is
depicted for quadaratic potential
\begin{equation}\label{quad}
V(\varphi)={1\over 2}m^2\varphi^2.
\end{equation}
This figure shows that, at the first stages after the slow-roll,
the amplitude of the scalar field drops down rapidly. Later,
during a phase of rapid oscillation of the field, the rate of
decrease of the amplitude becomes much less than the oscillation
frequency. To get an an insight about the solutions in this epoch,
we proceed in the same way as \cite{periodic} and consider the
power law potential (\ref{N1}), with an even integer $q$.

During the quasi-periodic evolution where the amplitude and the
frequency of oscillation are time dependent, the scalar field is
presented as \cite{periodic}
\begin{equation}\label{16}
\varphi(t)=\Phi(t)\cos\left(\int A(t)dt\right),
\end{equation}
where, the amplitude $\Phi(t)$ is given by
\begin{equation}\label{17}
V(\Phi(t))=\lambda \Phi^q(t)=\rho_{\varphi}(t).
\end{equation}
This equation means that the potential, when is evaluated at the
amplitude, gives all the energy. $A(t)$ is some function of time
which may be determined as follows: By taking a time derivative of
(\ref{16}), we easily obtain
\begin{equation}\label{a}
\sin\left(\int A(t)dt\right)= -{\dot{\varphi}\over
A\Phi}+{\dot{\Phi}\varphi\over A \Phi^2}.
\end{equation}
(\ref{a}), and (\ref{16}) result in
\begin{equation}\label{c}
A^2={\dot{\varphi}^2\left(1-{\varphi \over
\dot{\varphi}}{\dot{\Phi}\over \Phi}\right)^2\over
\Phi^2-\varphi^2}.
\end{equation}

The continuity equation and eqs. (\ref{4}) lead to:
\begin{equation}\label{18}
\dot{\rho_{\varphi}}=-3H[\dot{\varphi}^2(1+3wH^2-w\dot{H})-2wH\dot{\varphi}\ddot{\varphi}].
\end{equation}
By substituting $\ddot{\varphi}$ from the equation of motion
(\ref{3}) into the above equation we arrive at
\begin{equation}\label{d}
\dot{\rho_{\varphi}}=-3H[(9wH^2+3w\dot{H})\dot{\varphi}^2]+2\dot{\varphi}V_{,\varphi}.
\end{equation}
By using ${\dot{\Phi}\over \Phi}={1\over q}{\dot{\rho_{\varphi}}\over \rho_{\varphi}}$
derived from (\ref{17}), we deduce
\begin{equation}\label{e}
\left|{\varphi \over \dot{\varphi}}{\dot{\Phi}\over
\Phi}\right|=\left|\pm {6\varphi \over q
\rho_{\varphi}}\sqrt{{w(\rho_{\varphi}-V(\varphi))\over 2}}(3H^2+2\dot{H})+{2V(\varphi)\over
\rho_{\varphi}}\right|,
\end{equation}
where $+(-)$ corresponds to $\dot{\varphi}>0(<0)$.  This
expression is much less than unity, $\left|{\varphi \over
\dot{\varphi}}{\dot{\Phi}\over \Phi}\right|\ll 1$, whenever the
slow-roll is ceased and
\begin{equation}\label{20} \Phi\ll ({qM_P^2\over
6\sqrt{w\lambda}})^{2\over q+2},
\end{equation}
becomes valid. This is opposite to slow-roll conditions (see eqs.
(\ref{5},\ref{6}) and their subsequent discussion). In this regime
\begin{eqnarray}\label{N2}
A^2\approx {\dot{\varphi}^2 \over
\Phi^2-\varphi^2}&=&{2\left(\rho_{\varphi}-V(\varphi)\right)\over
9wH^2(\Phi^2-\varphi^2)}\nonumber \\
&=&{2M_P^2\left(\rho_{\varphi}-V(\varphi)\right)\over
3w\rho_{\varphi}(\Phi^2-\varphi^2)}.
\end{eqnarray}
$\left|{\dot{\Phi}\over \Phi}\right|\ll \left|{\dot{\varphi} \over
\varphi}\right|$ is the stage of rapid oscillation or high
frequency regime, i.e. when $\left|{\dot{\Phi}\over
\Phi}\right|\ll A$. In this epoch we have
\begin{equation}\label{22}
\left|{\dot{\Phi}\over \Phi}\right|=\left|{2\over q}{\dot{H}\over
H}\right|=\left|{1\over q}{\dot{\rho_{\varphi}}\over \rho_{\varphi}}\right|\ll A,
\end{equation}
showing that the amplitude, the Hubble parameter and the energy
density decrease slowly during one period of oscillation.

The time average of energy density over an oscillation cycle (from
$t$ to $t+T$) is
\begin{equation}
\left<\rho_{\varphi}(t)\right>:={\int_t^{t+T} \rho_{\varphi}(t')
dt' \over T},
\end{equation}
where $T$ is the period. As the amplitude decreases very slowly,
we deduce $ \left<\rho_{\varphi}(t)\right>=V(\Phi(t))$. This
result is in agreement with (\ref{17}), because from (\ref{22}),
it is clear that $\rho_{\varphi}$ changes insignificantly during a
period.

During the rapid oscillation of the scalar field, the parameter
$\gamma$, defined by $\gamma=1+{\left<P_{\varphi}\right>\over
\left<\rho_{\varphi}\right>}$, is determined as follows
\begin{equation}\label{23}
\gamma={\left<3wH^2\dot{\varphi}^2-{d\left({wH\dot{\varphi}^2}\right)\over
dt}\right>\over \left<\rho_{\varphi}\right>}.
\end{equation}
To obtain the above equation we have used (\ref{4}). Using
(\ref{4}), (\ref{23}) can be rewritten as
\begin{eqnarray}\label{24}
\gamma&=&{2\over 3}{\left<\rho_{\varphi}-V(\varphi)\right>\over
\left<\rho_{\varphi}\right>}\nonumber \\
&=&{2\over
3V(\Phi)}{\int_{-\Phi}^{\Phi}\sqrt{\rho_\varphi-V(\varphi)}d\varphi\over
\int_{-\Phi}^{\Phi}{d\varphi\over \sqrt{\rho_\varphi-V(\varphi)}}}\nonumber
\\
&=&{2\over 3}{\int_0^1 \sqrt{1-x^q}dx\over \int_0^1{1\over
\sqrt{1-x^q}}dx}={2q\over 3q+6}.
\end{eqnarray}
In the above computations, we have converted time integration to
$\varphi$ integration, and the variable change  $x={\varphi \over
\Phi}$ was applied. The same method for $w=0$, results in
$\gamma={2q\over q+2}$ \cite{periodic}. This constant is the
effective value of $\gamma$ in rapid oscillation era, i.e., in
this era  the scalar field behaves as a barotropic fluid. To
elucidate this point we proceed as \cite{periodic}. Taking time
average of (\ref{z}), and by using (\ref{24}), we obtain
\begin{equation}\label{time}
\dot{\left<\rho_{\varphi}\right>}+{2q\over q+2}H
\left<\rho_{\varphi}\right>=0.
\end{equation}

Using the definition of time average, we obtain  $\dot{\left<\rho_{\varphi}\right>}={\delta \rho_{\varphi}\over T}$,
where $\delta \rho_{\varphi}$ is the change of $\rho_{\varphi}$ over the
period $T$. Hence
\begin{equation}\label{time}
{\delta \rho_{\varphi}\over T}+{2q\over q+2}H
\left<\rho_{\varphi}\right>=0.
\end{equation}
In the high frequency regime  ${\delta \rho_{\varphi}\over
T}\approx \dot{\rho_{\varphi}}$, and we arrive at
\begin{equation}\label{25}
\dot{\rho_{\varphi}}+ {2q\over q+2}H\rho_{\varphi}=0.
\end{equation}

{\it{In the following, as in (\ref{25}), we will not use the
symbol $<>$ for the oscillating scalar field,  e.g., by
$\rho_\varphi$ we mean the time averaged of the scalar field
energy density in the sense explained above.}}

From eq. (\ref{17}) we get an
approximate equation for $\Phi$ evolution
\begin{equation}\label{26}
\dot{\Phi}\approx {-2\over q+2}H\Phi.
\end{equation}
From (\ref{26}) and (\ref{22}) it is obvious that $H\ll A$. This
means that the expansion rate is much less than the oscillation
frequency.  Effectively the scale factor is $a(t)\propto
t^{q+2\over q}$ and the Hubble parameter is given by $ H={q+2\over
q}{t^{-1}}$. $\rho$ satisfies
\begin{equation}\label{33}
{d(\rho_{\varphi} a^{3\gamma})\over dt}=0,
\end{equation}
corresponding to $\rho_{\varphi} \propto t^{-2}$ which is $q$ independent.

\section{Particle production}

In this part we try to study inflaton decay to ultra-relativistic
particles (radiation) in the rapid oscillatory phase. This decay
is due to the interaction between the inflaton and produced
particles. To study this decay, a phenomenological interaction
(see(\ref{34})),
\begin{equation} \label{phen}
Q=\Gamma \dot{\varphi}^2,
\end{equation}
where $\Gamma$ is a positive constant, was proposed by
\cite{stein}. Afterwards, in \cite{kof1},  a Lagrangian, including
bosonic and fermionic fields and their interactions with the
inflaton, was introduced, and it was shown that the effect of
particle production can be explained by adding a polarization
operator to the inflaton mass term. There was shown that, for
quadratic potential, the role of polarization operator may be
mimicked by the phenomenological friction term (\ref{phen}), in
high frequency regime. However, by taking under consideration the
back reaction of the quantum effects on the evolution of the
inflaton field \cite{kof2}, and also considering possible decays
of the inflaton to other particles \cite{alv}, (\ref{phen}) can no
more be deduced from a Lagrangian.

Other phenomenological models with temperature and field dependent
friction coefficient term, $Q=\Gamma (T,\varphi)\dot{\varphi}^2$,
have been also considered in the literature \cite{yok}.

However, as the nature of the inflaton and also primordial
produced particles are unknown, we have not yet an exact
expression for the form of interaction. In our study, we adopt the
widely used phenomenological interaction (\ref{phen}), which
reduces significantly computational complexity.

In the presence of the source term (\ref{phen}), the inflaton
evolution is given by
\begin{equation}\label{35}
3wH^2\ddot{\varphi}+3H(3wH^2+2w\dot{H})\dot{\varphi}+V'(\varphi)=-\Gamma\dot{\varphi}.
\end{equation}
In the following we consider the power law potential (\ref{N1}).
Using (\ref{4}) and with the same method used in (\ref{24}), one
can obtain
\begin{equation}\label{37a}
\left<9wH^2\dot{\varphi}^2\right>={2q\over q+2}\rho_{\varphi}.
\end{equation}
Inserting the above relation in the first equation of (\ref{34}),
and by replacing $P_{\varphi}+\rho_{\varphi}$ with its time
average over an oscillation cycle, we arrive at
\begin{equation}\label{38}
\dot{\rho}_{\varphi}+3H\gamma \rho_{\varphi}+{\gamma\Gamma\over
3wH^2}\rho_{\varphi}=0.
\end{equation}
We note again that in the above equation, all the values must be
regarded as the time averaged values over one oscillation. This
relation is only valid on large time with respect to the period of
fast oscillation. The second term (friction term) describes
dilution resulted from the universe expansion, while the third
term corresponds to particle production during the coherent
oscillation of the inflaton.

Comparing (\ref{38}) with the corresponding minimal model relation
\begin{equation}\label{N3}
\dot{\rho_{\varphi}}+3H\gamma_m\rho_{\varphi}+\gamma_m\Gamma\rho_{\varphi}=0,
\end{equation}
where $\gamma_m={2q\over q+2}$, shows that, in the limit $wH^2\gg
1$, the decrease of $\rho_{\varphi}$ (due to particle production)
and as we will see the reheating temperature are very less than
the corresponding values in the minimal model. Note (\ref{38}) is
true only for $wH^2\gg 1$, so (\ref{N3}) cannot be derived from
(\ref{38}) by simply setting $w=0$.

\begin{figure}[h!]
\centering\epsfig{file=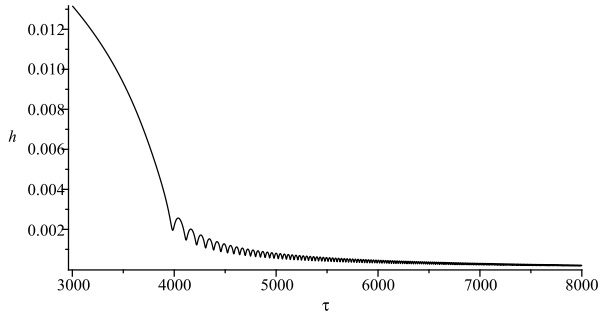,width=8cm,angle=0}
\caption{$h:={H\over m}$ in terms of dimensionless time $\tau=mt$,
for $\{wm^2=10^{8}, {\Gamma\over m}= 10^{-2}, \}, $ with initial
conditions $\{\phi(1)=0.056$, $\dot{\phi}(1)=0\}$, for the quadratic potential.} \label{fig2}
\end{figure}

The solution of (\ref{38}) is
\begin{equation}\label{39}
\rho_{\varphi}\propto a^{-3\gamma}\exp\left[{-\Gamma \gamma \over
3w}\int_{t_{osc.}}^tH^{-2}(t')dt'\right],
\end{equation}
where $t_{osc.}$ is the time when the oscillation commences. In
the beginning of particle production, the universe is dominated by
the scalar field, and we assume $\Gamma \ll wH^3$ , but $H$ is
decreasing and this approximation ceases to be valid later, when
the second and the third term in (\ref{38}) acquire the same order
of magnitude (we will denote this time by $t_{rh}$). So, with our
assumptions, and in the scalar field dominated era it is safe to
use the approximation $H={2\over 3\gamma t}$ until $t=t_{rh}$. The
behavior of the Hubble parameter, during rapid oscillation, can
also be investigated via numerical method. E.g. in
fig.(\ref{fig2}), $H$ is plotted for the quadratic potential
(\ref{quad}) and by using Friedmann and continuity equations.
So let us write (\ref{39}) as
\begin{equation}\label{40}
\rho_{\varphi}=\Xi t^{-2}\exp(-{\Gamma \gamma^3\over 4w}t^3),
\end{equation}
where $\Xi=t_{osc.}^2\rho_{osc.}e^{{\Gamma\gamma^3\over 4w}
t_{osc.}^3}$ and $\rho_{osc.}=\rho(t_{osc.})$. The decrease of
$\rho_{\varphi}$ is due to particle production which is encoded in
the exponential term and also due to the term $t^{-2}$
corresponding to dilution via the universe expansion. The dilution
term is independent of $q$ as was explained after eq. (\ref{33}).
For larger values of ${\Gamma\over wM_P^3}$ the decay rate is
faster.

Comparing this result with what was obtained before
for \{$w=0$,
$\gamma=0.5$\} model \cite{kolb},
\begin{equation}\label{41}
\rho_{\varphi}=\rho_{osc.}({t_{osc}\over
t})^2\exp[-\Gamma(t-t_{osc.})],
\end{equation}
shows that the decrease rate of $\rho_{\varphi}$, due to the
expansion of the universe, is the same. But in the presence of
$w$, the rate of particle production is decreased. We note again
that our result (\ref{40}) is true only for $wH^2\gg 1$, so we
cannot obtain (\ref{41}) from (\ref{40}) by setting $w=0$.

\begin{figure}[h]
\centering\epsfig{file=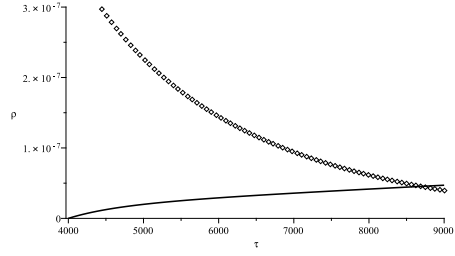,width=6cm} \caption{dimensionless
energy density ${\rho_{\varphi}\over m^2 M_P^2}$ (points) and
${\rho_{\mathcal{R}}\over m^2M_P^2}$ (line) in terms of
dimensionless time $\tau=mt$, for $\{wm^2=10^{8}, {\Gamma\over m}=
10^{-2}, \}$ with $\rho_{osc.}=8.3\times 10^{-8}m^2M_P^2$, for the
quadratic potential} \label{fig3}
\end{figure}

The radiation satisfies
\begin{equation}\label{42}
\dot{\rho}_{\mathcal{R}}+4H\rho_{\mathcal{R}}=\Gamma\gamma{\rho_{\varphi}\over
3wH^2}.
\end{equation}
To study the evolution of $\rho_{\mathcal{R}}$ during scalar field
dominated epoch, i.e., when  $H^2\simeq {1\over
3M_P}\rho_{\varphi}$ we write (\ref{42}) in the form
\begin{equation}\label{43}
\dot{\rho}_{\mathcal{R}}+4H\rho_{\mathcal{R}}={\Gamma \gamma
M_P^2\over w},
\end{equation}
whose approximate solution is given by
\begin{equation}\label{444}
{\rho}_{\mathcal{R}}={3\Gamma \gamma^2 M_P^2\over
(8+3\gamma)w}t[1-({t_{osc.}\over t})^{{8+3\gamma}\over 3\gamma}].
\end{equation}
To derive the above equation we have assumed that after the
slow-roll, the universe was cold:
$\rho_{\mathcal{R}}(t=t_{osc.})=0$. In fig. (\ref{fig3}),
${\rho}_{\mathcal{R}}$ is plotted for the quadratic potential
(\ref{quad}), showing that the radiation density increases
monotonically in the rapid oscillation phase. However this
behavior is valid only for $\rho_{\varphi} \gtrsim
\rho_{\mathcal{R}}$, and in the radiation dominated era we expect
that $\rho_{\mathcal{R}}$ decreases.

\begin{figure}[h]
\centering\epsfig{file=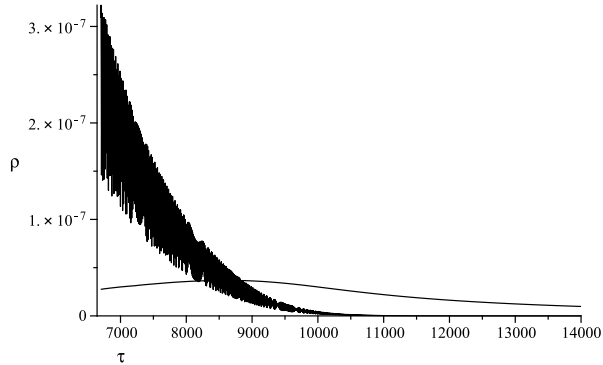,width=8cm,angle=0}
\caption{dimensionless energy density $\rho_{\varphi}\over m^2
M_P^2$ (initially upper graph) and ${\rho_{\mathcal{R}}\over
m^2M_P^2}$ in terms of dimensionless time $\tau=mt$, for
$\{wm^2=10^{8}, {\Gamma\over m}= 10^{-2}, \}, $ with initial
conditions $\{\phi(\tau=1)=0.056$, $\dot{\phi}(\tau=1)=0\}$, for the quadratic potential. }
\label{fig4}
\end{figure}

Based on Friedmann and continuity equations, the behavior of
${\rho}_{\mathcal{R}}$ is also depicted in fig.(\ref{fig4}), via
numerical method for the quadratic potential (\ref{quad}). This
figure shows that the increase of $\rho_{\mathcal{R}}$ continues
until ${\rho}_{\mathcal{R}}\simeq {\rho}_{\varphi}$. After
$t=t_{rh}$, i.e. from the beginning of radiation dominated era,
${\rho}_{\mathcal{R}}$ begins to decrease.

Relativistic particles interact quickly (with respect to the
expansion rate of the universe) with each other to become in a
thermal equilibrium characterized by temperature $T_r$, given by
\begin{equation}\label{45}
\rho_{\mathcal{R}}={\pi^2\over 30}g_{*}T_r^4,
\end{equation}
where $g_{*}$ is the total number of effectively massless degrees
of freedom. If $T_r$ is greater than the electroweak scale
$T_r>300GeV$ then $g>106.75$ \cite{milc}. The reheating time,
$t_{rh}$, is given by
\begin{equation}\label{N4}
\Gamma \simeq 9wH^3(t_{rh}),
\end{equation}
i.e. , when the second and third terms of (\ref{38}) acquire a
same order of magnitude, as it was mentioned before. At this order
of time, ${\rho_{\mathcal{R}}(t_{rh})}$ has also the same order of
magnitude as $\rho_{\varphi}(t_{rh})$. At the reheating time we
have $H\simeq \left({\Gamma\over 9w}\right)^{1\over 3}$, so the
use high friction condition (\ref{f}) during the reheating era is
safe only when
\begin{equation}\label{N5}
w\Gamma^2\gg 1.
\end{equation}

For $t \gtrsim t_{rh}$, almost all the energy of the inflaton is
transferred to newly produced particles and the universe becomes
radiation dominated.

\begin{figure}[h!]
\centering\epsfig{file=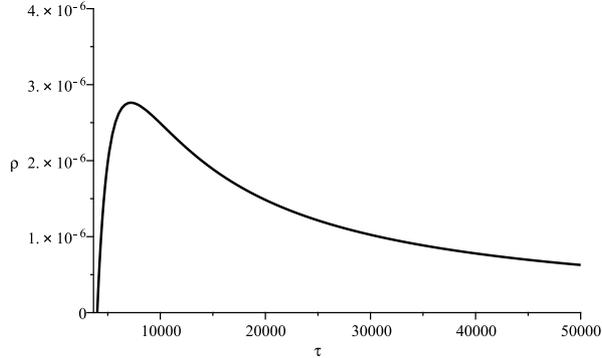,width=8cm,angle=0}
\caption{${\rho_{\mathcal{R}}\over m^2M_P^2}$, in terms of
dimensionless time $\tau=mt$, for $w=0$, ${\Gamma\over m}=10^{-2}$
with $\rho_{osc.}=8.3\times 10^{-8}m^2M_P^2$, for the quadratic
potential.}\label{fig5}
\end{figure}

Now, we can estimate the reheating temperature defined by
$T_{rh}=T(t_{rh})$. From (\ref{444}), and ${t_{osc.}\over
t_{rh}}\ll 1$, we obtain
\begin{equation}\label{47}
T_{rh}\simeq 1.89\left({\gamma\over 8+3\gamma}\right)^{1\over
4}g_{*}^{-{1\over 4}}M_P^{1\over 2}({\Gamma\over w})^{1\over 6}.
\end{equation}
Note that the above equation could also be obtained by using $H^2
\simeq {1\over 3M_P^2}\rho_{\mathcal{R}}$, and (\ref{45}). This
temperature is specified only by parameters of the system and is
independent of initial conditions. In the absence of non-minimal
derivative coupling, i.e. for $w=0$, and for $\gamma=0.5$, the
radiation energy density is obtained as \cite{kolb}
\begin{equation}\label{46}
\rho_{\mathcal{R}}={M_P^2\Gamma\over 10\pi t}[1-({t_{osc.}\over
t})^{5\over 3}].
\end{equation}
$\rho_{\mathcal{R}}$ increases rapidly from $\rho_{\mathcal{R}}=0$
to its maximum value and then decreases again, so the maximum
temperature is occurred in the beginning of $\varphi$ oscillation
before reheating (see fig.(\ref{fig5}), depicted for the potential
(\ref{quad})).
This is in contrast to our model, where as it can be seen from
(\ref{444}), $\rho_{\mathcal{R}}$ increases continuously in
$\varphi$ oscillation epoch until
$\rho_{\mathcal{R}}\simeq\rho_{\varphi}$. In $w=0$ model, reheating
temperature is determined by \cite{kolb}
\begin{equation}\label{1000}
T_{rh} (w=0)\simeq 1.2g_{*}^{-{1\over 4}}M_P^{1\over
2}\Gamma^{1\over 2}.
\end{equation}
Therefore $ T_{rh}(wH^2\gg 1)\ll T_{rh} (w=0)$, provided that a
same $\Gamma$ is taken into account for both theories. Note that
as the reheating process is realized after inflation, $T_{rh}$
must be below the GUT scale: $T_{rh}< 10^{16} GeV$.

Based on astrophysical data, we are able to estimate the relation
of reheating temperature and the number of e-folds. To do so, we
proceed as follows: Consider a length scale $l$ which at time
$t=t_{*}$, in the inflation era, left the Hubble radius:
\begin{equation}\label{N6}
l={a_0\over a(t_{*})}{1\over H(t_{*})}
\end{equation}
The number of e-folds from $t_{*}$ to the end of the slow-roll
inflation, denoted by subscript $e$,  may be expressed as
\begin{equation}\label{N7}
\mathcal{N}_{*}=\ln \left({a_e\over a_{*}}\right)=\ln
\left({a_e\over a_{rh}}{a_{rh}\over a_{eq}}{a_{eq}\over
a_0}{a_0H_0\over a_{*}H_{*}}{H_{*}\over H_0}\right),
\end{equation}
where $0$, $rh$, and $eq$ subscripts denote present, reheating,
and matter-radiation equality densities epochs respectively. As we
have seen, during rapid oscillation (from $t_e$ until $t_{rh}$),
the universe is dominated by a scalar field whose the effective
equation of state parameter is given by $\gamma-1$, whence
\cite{lid}
\begin{equation}\label{N8}
\mathcal{N}_{*}=62+\ln \left({a_0H_0\over a_{*}H_{*}}\right)+\ln
\left({V_{*}^{1\over 4}\over 10^{16}GeV}\right)+{1\over 4}\ln
\left({V_{*}\over V_e}\right)-\left({{1\over 3\gamma}-{1\over
4}}\right)\ln \left({V_e\over \rho_{rh}}\right).
\end{equation}
The right hand side expressions are determined as follows: By
setting $a_0=1$, and adopting the WMAP (pivot) scale $l={1\over
k}=500 Mpc$ \cite{Koma}, we arrive at
\begin{equation}\label{N9}
\ln \left({a_0H_0\over a_{*}H_{*}}\right)=\ln\left({H_0\over
k}\right)=\ln\left({0.7\over 6}\right).
\end{equation}
The present Hubble parameter is $H_0={7\over 30000}Mpc^{-1}$
\cite{Koma}.  To obtain $V_{*}$, we consider the scalar
perturbations. The spectral index $n_s$ is \cite{hf}
\begin{equation}\label{N10}
n_s-1={M_P^2\over wH^2}\left[{2\over 3}{V''(\varphi)\over
V(\varphi)}-{4\over 3}{V'(\varphi)\over V(\varphi)^2}\right].
\end{equation}
The quantities in the right hand side must computed at the time of
horizon crossing $c_sk=aH$ during slow-roll inflation, where $k$
is the comoving wavenumber and $c_s$ is the sound speed of scalar
perturbation. In the limit $wH^2\gg 1$ we have $c_s\simeq 1$
\cite{cs}, and for the potential (\ref{N1}),  we deduce
\begin{equation}\label{N11}
V_{*}=\lambda^{2\over q+2}\left({2M_P^4q(q+2)\over
w(1-n_s)}\right)^{q\over q+2}.
\end{equation}
Based on WMAP data \cite{Koma},
\begin{equation}\label{N12}
n_s=0.968 \pm0.012.
\end{equation}
From $t_{*}$ until $t_{e}$, the slow-roll approximation is valid,
hence $V_{e}$ may be estimated with the help of (\ref{Y1}), as:
\begin{equation}\label{N13}
V_{e}=\lambda^{2\over q+2}\left(q^2M_P^4\over w\right)^{q\over
q+2}.
\end{equation}
$\rho_{rh}$ is specified by (\ref{444}) or equivalently by (\ref{45}) and (\ref{47}):
\begin{equation}\label{N14}
\rho_{rh}={4.16 M_P^2\gamma\over {8+3\gamma}} \left({\Gamma\over
w}\right)^{2\over 3}.
\end{equation}
In the slow-roll regime, as expected, we have $V^{1\over
4}_{*}\simeq V^{1\over 4}_{e}$.  $V_{*}^{1\over 4}\lesssim
10^{16}GeV$ also holds \cite{lid} ($10^{16}GeV$ is the GUT scale).
The reheating is provided by the inflaton energy, hence
$\rho_{rh}\lesssim V_{e}$. So finally we expect to have
$\mathcal{N}_{*}< 60$. Only for a prompt reheating, $N\approx 60$
may be possible.

As an example, for the quadratic potential $V(\varphi)={1\over
2}m^2\varphi^2$ where
\begin{eqnarray}\label{N15}
V_e&=&{\sqrt{2} m M_P^2\over  \sqrt{w}}\nonumber \\
V_{*}&=&{\sqrt{6}mM_P^2\over \sqrt{(1-n_s)w}}\nonumber \\
\rho_{rh}&=&0.15 M_P^2 \left({\Gamma\over w}\right)^{2\over 3}.
\end{eqnarray}
$\mathcal{N}_{*}$ depends on the parameters of the model, i.e.
$\Gamma$, $w$ and $m$, e.g. choosing $wm^2\simeq 10^8$,
${\Gamma\over m}\simeq 10^{-2}$ (in agreement with (\ref{N5})),
and setting $m=10^{-6}M_P$ \cite{inflaton1}, we find
$\mathcal{N}_{*}=43.41$.

\section{Conclusion}
As a summary, we briefly discussed inflation in the framework of a
non-minimal derivative coupling model proposed in \cite{germani},
and then studied a gap in the literature: the reheating process in
this framework. We investigated inflaton evolution in
quasi-periodic oscillation at the end of slow-roll. We allowed the
scalar field to decay to ultra-relativistic particles (radiation)
via a phenomenological source term. We obtained the reheating
temperature which was independent of initial conditions.

We showed that the energy density of radiation, during oscillatory
era and when it is smaller than the energy density of inflaton,
increases monotonically. This behavior is in contrast to ordinary
inflation theory where the maximum temperature occurs before
reheating. We confirmed our results via numerical methods.

\end{document}